\documentstyle[12pt]{article}
\begin{document}
\begin{center}
{\Large
Supercooled Liquids and Glasses\index{glasses}\index{supercooled
liquids}\footnote{Lectures presented at {\it Soft and Fragile Matter,
Nonequilibrium Dynamics, Metastability and Flow} University of St.
Andrews, 8 July -- 22 July, 1999}
}
\end{center}
\vspace*{10mm}

\begin{center}
{\large
Walter Kob\\
}
Institute of Physics, Johannes-Gutenberg University, D-55099 Mainz, Germany
\end{center}

\section{Introduction}
\label{sec1}

Glasses are materials that are much more common in our daily life than
one might naively expect. Apart from the obvious (inorganic) glasses,
such as wine glasses, bottles and windows, we also have the organic
(often polymeric) glasses, such as most plastic materials (bags, coatings,
etc.). In the last few years also metallic glasses\index{metallic glasses}
have come to our daily (St.  Andrews) life since they are used, apart
from many other applications, in the head of golf clubs.  In view of
this widespread use of these materials it might be a bit surprising
to learn that glasses are not very well understood from a microscopic
point of view and that even today very basic questions such as ``What
is the difference between a liquid and a glass?'' cannot be answered in
a satisfactory way. In the present lecture notes we will discuss some of
the typical properties of supercooled liquids and glasses and theoretical
approaches that have been used to describe them. Since unfortunately
it is not possible to review here all the experiments on glasses
and theoretical models to explain them we will discuss here only some of
the most basic issues and refer the reader who wants to learn more about
this subject to other review articles and textbooks~\cite{glass_reviews}.

In the following section we will review some of the basic phenomena that
are found in supercooled liquids and glasses. Subsequently we will discuss
the theoretical approaches to describe the dynamics of these systems,
notably the so-called mode-coupling theory of the glass transition. This
will be followed by the presentation of results of computer simulations
to check to what extend this theory is reliable. These results are
concerned with the {\it equilibrium} dynamics. If the temperature of the
supercooled liquid is decreased below a certain value, the system is no
longer able to equilibrate on the time scale of the experiments, i.e. it
undergoes a glass transition. Despite the low temperatures the system
still shows a very interesting dynamics the nature of which is today
still quite unclear. Therefore we will present in the final part of these
lecture notes a brief discussion of this dynamics and its implication
for the (potential) connection of structural glasses with spin glasses.

\section{Supercooled Liquids and the Glass Transition}\index{glass transition}
\label{sec2}

In this section we will discuss some of the properties of supercooled
liquids and some of the phenomena of the glass transition.

\begin{figure}[hbt]
\vspace*{100mm}
\includegraphics{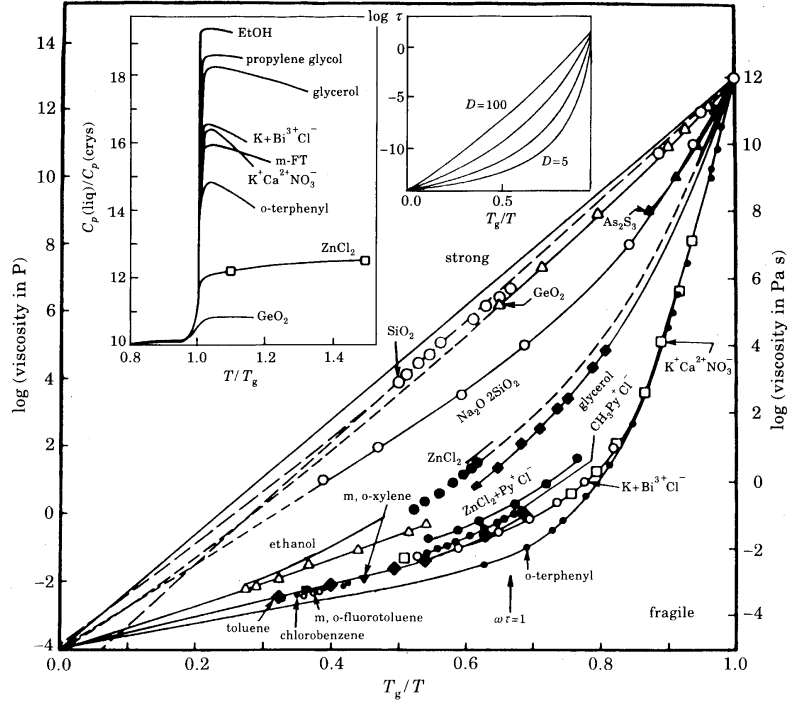}
\caption{Main figure: Viscosity of different glass-formers as a
function of $T_g/T$, where $T_g$ is the glass transition temperature.
Left inset: Temperature dependence of the specific heat, normalized
to its value for the crystall, for different glass-formers.  From
reference~\protect\cite{angell94}, with permission.}
\label{fig1}
\end{figure}

If a liquid is cooled from high temperatures below its melting point
$T_m$ one expects it to crystallize at $T_m$. However, since the
crystallization process takes some time (critical nuclei have to be
formed and have to grow) it is possible to supercool most liquids, i.e.
they remain liquid-like even below $T_m$. Some liquids can be kept in
this metastable state for a long time and thus it becomes possible
to investigate their properties experimentally. For reasons that will
become clear below, such liquids are called good glass-formers. It is
found that with decreasing temperature the viscosity\index{viscosity}
$\eta$ of these systems increases by many orders of magnitude. In order
to discuss this strong temperature dependence it is useful to define
the so-called glass transition temperature $T_g$\index{glass transition
temperature} by requiring that at
$T_g$ the viscosity is $10^{12}$ Pa s, which corresponds {\it roughly}
to a relaxation time of 100 seconds (reminder: water at room temperature
has a viscosity around $10^{-3}$ Pa s). In figure~\ref{fig1} we show the
temperature dependence of $\log(\eta)$ for a variety of glass-formers
as a function of $T/T_g$. From that plot we see that the viscosity does
indeed increase dramatically when temperature is decreased. Furthermore
we recognize that this temperature dependence depends on the material
in that there are substances in which $\eta(T)$ is very close to an
Arrhenius law\index{Arrhenius law}, i.e. are almost straight lines, and
other substances in which a pronounced bend in $\eta(T)$ is found. In
order to distinguish these different temperature dependencies Angell
are coined the terms ``strong'' and ``fragile'' glass-formers\index{strong glass
former}\index{fragile glass former} for the
former and latter, respectively~\cite{angell85}.

The strong temperature dependence which is found in $\eta(T)$ is not a
unique feature of the viscosity. If other transport quantities, such as
the diffusion constant\index{diffusion constant}, or relaxation times are
measured, it is found that they show a similar temperature dependence
as the viscosity. On the other hand if thermodynamic quantities, such
as the specific heat, or structural quantities, such as density or
the structure factor, are measured, they show only a relatively mild
dependency in the same temperature interval, i.e. they vary between 10\%
and a factor of 2-3.

Equipped with these experimental facts one can now ask the main question
of glass physics: What is the reason for the dramatic slowing down of the
dynamics of supercooled liquids without an apparent singular behavior of
the static quantities? Although this question seems to be a very simple
one it has not been possible up to now to find a completely satisfying
answer to it. One obvious response is to postulate the existence of a
second order phase transition\index{phase transition} at temperatures
below $T_g$. Then the slowing down could be explained as the usual
critical slowing down observed at the critical point. Although such
an explanation is from a theoretical point of view very appealing it
suffers one big drawback, namely that so far it has not been possible to
identify an order parameter which characterizes this phase transition
or a characteristic length scale which diverges. Thus despite the nice
theoretical concept the phase transition idea is not able to provide a
satisfactory explanation for the slowing down of the dynamics.

Things look much better for a different theoretical approach, the
so-called mode-coupling theory (MCT) \index{mode-coupling theory} of the
glass transition, which we will discuss in mode detail below. This theory
is indeed able to make qualitative and quantitative prediction for the
time and temperature dependence of various quantities and experiments
and computer simulations have shown that many of these predictions are
true~\cite{gotze99}. However, before we discuss the predictions of MCT we
return to the temperature dependence of the viscosity or the relaxation
times. From figure~\ref{fig1} it is clear that for every material there
will be a temperature at which the relaxation time of the system will
exceed by far any experimental time scale. This means that it will not
be possible to probe the {\it equilibrium} behavior of the system below
this temperature. If the system is continously with a given cooling
rate from a high temperature to low temperatures there will exist a
temperature $T_g'$ at which the typical relaxation time of the system
is comparable to the inverse of the cooling rate. Hence in the vicinity
of this temperature the system will fall out of equilibrium and become
a glass. (Note that the other glass transition temperature that we have
introduced above, $T_g$, is basically the value of $T_g'$ if one assumes
that the relaxation time is on the order of 100 s.) This glass transition
is accompanied by the freezing of those degrees of freedom which lead to a
relaxation of the system, such as the motion of the particles beyond the
nearest neighbor distance. Since below $T_g'$ these degrees of freedom
are no longer able to take up energy the specific heat shows a drop at
$T_g'$, as it can be seen in the left inset of figure~\ref{fig1}. Note
that empirically it is found that the fragile glass-formers show a large
drop in the specific heat whereas the strong glass-formers show only a
small one. Note, however, that this correlation is just an empirical one
(and it does not hold strictly) and apart from hand-waving arguments it
is not understood from a theoretical point of view. The same is also true
for the distinction between strong and fragile glass-formers. So far it
is not clear what the essential features in a Hamiltonian are that make
the system strong or fragile, i.e. is this the range of the interaction,
the coordination number, etc.

At the beginning of this section we mentioned that the dynamics of
glass-forming liquids becomes slow when they are cooled below the
melting temperature. However, it is not a necessary condition for a
slow dynamics that the temperature is below $T_m$.  E.g. silica has a
melting temperature around 2000K and a glass transition temperature around
1450K~\cite{mazurin83}. From Figure~\ref{fig1} it becomes obvious that at
$T_m=T_g/0.725$ the viscosity is already on the order of $10^7$Pa s! Thus
it is clear that slow dynamics has nothing to do with the system being
supercooled, or in other words: for the glass transition the melting
temperature is a completely irrelevant quantity. Despite this fact we
will in the following continue to talk about ``supercooled'' liquids,
following the usual (imprecise) usage of this term.

We now turn our attention to the MCT~\index{mode-coupling theory}, 
the theory we have briefly
mentioned earlier. Here we will give only a very sketchy idea about this
theory and refer the reader who wants to learn more about it to the
various review articles on MCT~\cite{gotze99,mct_reviews}. In the MCT
the quantities of interest are the correlation functions between the
density fluctuations of the particles. If we denote by ${\bf r}_j(t)$
the position of the particle $j$ at time $t$ the density fluctuations\index{density
fluctuations}
are given by~\cite{hansenmcdonald86}
\begin{equation}
\delta \rho({\bf q},t)=\sum_{j=1}^N \exp(i {\bf q} \cdot {\bf r}_j(t))\quad,
\label{eq1}
\end{equation}
where ${\bf q}$ is the wave-vector. From this observable one can calculate
the so-called intermediate scattering function $F(q,t)$\index{intermediate
scattering function} which is given by

\begin{equation}
F(q,t)=\frac{1}{N}\langle \delta\rho(-{\bf q},t)\delta \rho({\bf q},0) \rangle
\quad.
\label{eq2}
\end{equation}

Here the angular brackets stand for the thermodynamic average. The
relevance of the function $F(q,t)$ is given by the fact that it can be
directly measured in neutron and light scattering experiments. From a
theoretical point of view this correlation function is important since
many theoretical descriptions of (non-supercooled) liquids are based on
it, or its time and space Fourier transforms~\cite{hansenmcdonald86}.

Using the Mori-Zwanzig projection operator
formalism~\cite{hansenmcdonald86} \index{Mori-Zwanzig formalism} it 
is now possible to derive {\it exact}
equations of motion for the $F(q,t)$. These are of the form
\begin{equation}
\ddot{F}(q,t)+\Omega^2(q)F(q,t)+\int_0^t d\tau M(q,\tau)\dot{F}(t-\tau)=0.
\label{eq3}
\end{equation}
Here $\Omega^2(q)$ is given by $q^2k_BT/mS(q)$, where $m$ is the
mass of the particles and $S(q)$ is the static structure factor,
i.e. $S(q)=F(q,0)$\index{structure factor}. The function $M(q,\tau)$ is called 
the memory function\index{memory function}
and formally exact expression exist for it. However, because of their
complexity, these formal expressions are basically useless for a real
calculation and thus in MCT one approximates $M(q,\tau)$ by a quadratic
form of the density correlators. In particular it is found that $M(q,t)$
is given by
\begin{equation}
M(q,t)=\frac{1}{2 (2\pi)^3}\int d{\bf k} 
V^2(q,k,|{\bf q}-{\bf k}|)F(k,t) F(|{\bf q}-{\bf k}|,t)
\label{eq4}
\end{equation}
where the vertex $V^2$ is given by
\begin{equation}
V^2(q,k,|{\bf q}-{\bf k}|)=\frac{n}{q^2}
\left(\frac{{\bf q}}{q}\left[ {\bf k}c(k)+
({\bf q}-{\bf k}) c(|{\bf q}-{\bf k}|) \right] \right)
\label{eq5}
\end{equation}
and the so-called direct correlation function $c(k)$ can be expressed
via the structure factor by $(S(k)-1)/nS(k)$, where $n$ is the
particle density.  Thus we see that within MCT the static structure
factor determines the vertex $V^2$, which in turn determines the memory
function for the time dependent correlation function. Or in other words:
The statics determine the dynamics.

Note that similar equations of motion as the one for $F(q,t)$ exist for the
incoherent intermediate scattering function, $F_s(q,t)$. This correlation
function is given by
\begin{equation}
F_s(q,t)=\frac{1}{N}\left\langle \sum_j^N \exp(i{\bf q}\cdot ({\bf r}_j(t)-{\bf
r}_j(0)) \right\rangle
\quad,
\label{eq6}
\end{equation}
i.e. it is just the self (or diagonal) part of $F(q,t)$. Also this time
correlation function is important since it can be measured in scattering
experiments. 

Instead of making at this point a detailed discussion of the properties
of the solutions of these MCT equations we will postpone this discussion
to section~\ref{sec4} where we will make a detailed comparison of the
prediction of MCT with the results of computer simulations. The only
thing that we mention already now is that it has been shown that {\it at
long times} there are two types of solution of the MCT equations. The
first one is the solution $\lim_{t\to \infty} F(q,t)=0$. This solution
is the only one at high temperature and it corresponds to the physical
situation that the system is ergodic, i.e. all time correlation functions
decay to zero. (Note that temperature enters though the temperature
dependence of the static structure factor $S(q)$). The second solution
has the property that $\lim_{t\to \infty} F(q,t)>0$ and it occurs only
below a critical temperature $T_c$\index{critical temperature}. Since
in this case the correlation functions do not decay to zero even at
long times, the system is no longer ergodic, i.e. it is a glass. Thus
within the MCT the system undergoes a glass transition at $T_c$. MCT now
makes an asymptotic expansion of the dynamics around this critical point,
i.e. it treats the quantity $\epsilon=(T_c-T)/T_c$ as a small parameter.
Hence all the predictions of the theory are, strictly speaking, only valid
very close to $T_c$ and it is difficult to say a priori how far away from
$T_c$ they are still useful. However, our experience of analyzing data
has shown that the theory can be used for values $\epsilon$ as large as
0.5 or so~\cite{gotze99}, thus with respect to this the situation seems
to be much better than the case of critical phenomena.

\section{On Computer Simulations}\index{computer simulations}
\label{sec3}

In the last few decades computer simulations have been shown to be a
very powerful tool to gain insight into the behavior of statistical
mechanic systems and thus can be considered to be a very useful
addition to experiments and analytical calculations. Present days
computer codes for such simulations are usually quite complex and thus
we are not going to discuss the various tricks used in such simulations
but refer the reader to some text books and the lecture of K. Kremer in
this school~\cite{simul_books,kremer99}.

Simulations of supercooled liquids and glasses pose special problems
for computer simulations since at low temperatures the relaxation times
are large, see the previous section, and thus the simulations have to
be done for many (microscopically small!) time steps. Fortunately it is
usually not necessary to use very large system sizes, a few hundred to
a few thousand particles are adequate for most cases, and thus all the
computer resources are spend to simulate the system over a time span
which is as large as possible. Therefore present day state of the art
calculations extend over 10-100 million time steps which corresponds
to several month up to several years of CPU time on a top of the line
processor. Note that despite this effort the length of such a runs
corresponds to only about $10^{-7}$ seconds, since each time step is on
the order of $10^{-15}$ seconds. However, it should be noted that the
time window of these simulations, i.e. 7-8 decades, exceeds the one of
most experimental techniques, such as neutron or light scattering. A
more extensive discussion of advantages and disadvantages of computer
simulations of supercooled liquids and glasses and references to the
original literature can be found in Refs.~\cite{glass_sim_rev}.

We now discuss some of the details of the simulations whose results
will be discussed in the next few sections. As mentioned in the previous
paragraph the main issue of computer simulations of supercooled liquids
is to investigate the system at a temperature which is as close as
possible to the glass transition temperature, i.e. in that temperature
range where the relaxation times of the system are large. Therefore
it is advisable to use a system that can be simulated as efficiently
as possible. Hence many investigations have been done for so-called
``simple liquids''\index{simple liquids}, i.e. systems in which the
interaction between the particles is isotropic and short ranged. One
possible example for such a system is a one-component Lennard-Jones
liquid. The main drawback of this system is that it is prone to
crystallization, i.e. something which in this business has to be avoided
at all costs. Therefore it has become quite popular to study {\it binary}
liquids, since the additional complexity of the system is sufficient to
prevent crystallization, at least on the time scale accessible to todays
computer simulations.

The system we study is hence a binary Lennard-Jones\index{Lennard-Jones} 
liquid and in
the following we will call the two species of particles ``type A''
and ``type B'' particles. The interaction between two particles of type
$\alpha$ and $\beta$, $\alpha, \beta \in \{{\rm A,B}\}$, is thus given by:
$V_{\alpha\beta}= 4\epsilon_{\alpha\beta}[(\sigma_{\alpha\beta}/r)^{12}-
(\sigma_{\alpha\beta}/r)^6]$. The values of the parameters
$\epsilon_{\alpha\beta}$ and $\sigma_{\alpha\beta}$
are given by $\epsilon_{AA}=1.0$, $\sigma_{AA}=1.0$,
$\epsilon_{AB}=1.5$, $\sigma_{AB}=0.8$, $\epsilon_{BB}=0.5$, and
$\sigma_{BB}=0.88$. This potential is truncated and shifted at a distance
$\sigma_{\alpha\beta}$. In the following we will use $\sigma_{\rm AA}$
and $\epsilon_{\rm AA}$ as the unit of length and energy, respectively
(setting the Boltzmann constant $k_{\rm B}=1.0$). Time will be measured
in units of $\sqrt{m\sigma_{\rm AA}^2/48\epsilon_{\rm AA}}$, where $m$
is the mass of the particles.

In the following we will study two types of dynamics for this system:
A Newtonian dynamics\index{Newtonian dynamics} (ND) and a stochastic
dynamics\index{stochastic dynamics}\index{Brownian dynamics} (SD). The
reason for investigating the ND is that this is a realistic dynamics for
an atomic liquid. Thus it is possible to study at low temperatures the
interaction of the phonons with the relaxation dynamics of the system.
On the other hand the SD is a good model for a colloidal suspension in
which the particles are constantly hit by the (much smaller) particles of
the bath. In such systems the phonons are strongly damped and thus such
a dynamics is one way to ``turn off'' the phonons. Hence, by comparing
the results of the two types of dynamics it becomes possible to find
out which part of the dynamics is universal, i.e. does not depend on
the microscopic dynamics, and which part is non-universal.

In both types of simulations the number of A and B particles were 800
and 200, respectively. The volume of the simulation box was kept
constant at a value of $(9.4)^3$, which corresponds to a particle
density of around 1.2.  The temperatures used were 5.0, 4.0, 3.0, 2.0,
1.0, 0.8, 0.6, 0.55, 0.5, 0.475, 0.466, 0.456, and 0.446.  At each
temperature the system was thoroughly equilibrated for a time span
which significantly exceeded the typical relaxation times of the system
at this temperature. At the lowest temperatures this took up to 40 million
time steps. As we will see, the relaxation times for the SD are, at low
temperatures, significantly longer than the ones for the ND. Therefore
we used in all cases the ND to equilibrate the sample and used the SD
only for the production runs. For the ND we used at low temperatures a
time step of 0.02, whereas for the SD a smaller one, 0.008, was needed
in order to avoid systematic errors in the equilibrium quantities. In
order to improve the statistics of the results we averaged over eight
independent runs.

\section{The Equilibrium Relaxation Dynamics}
\label{sec4}

In this section we will discuss the relaxation dynamics of the system {\it
in equilibrium}. The main emphasis will be to find out to what extend
this dynamics depends on the microscopic dynamics and which aspects of
it can be understood within the framework of the mode-coupling theory.

\begin{figure}[hbt]
\vspace*{90mm}
\includegraphics{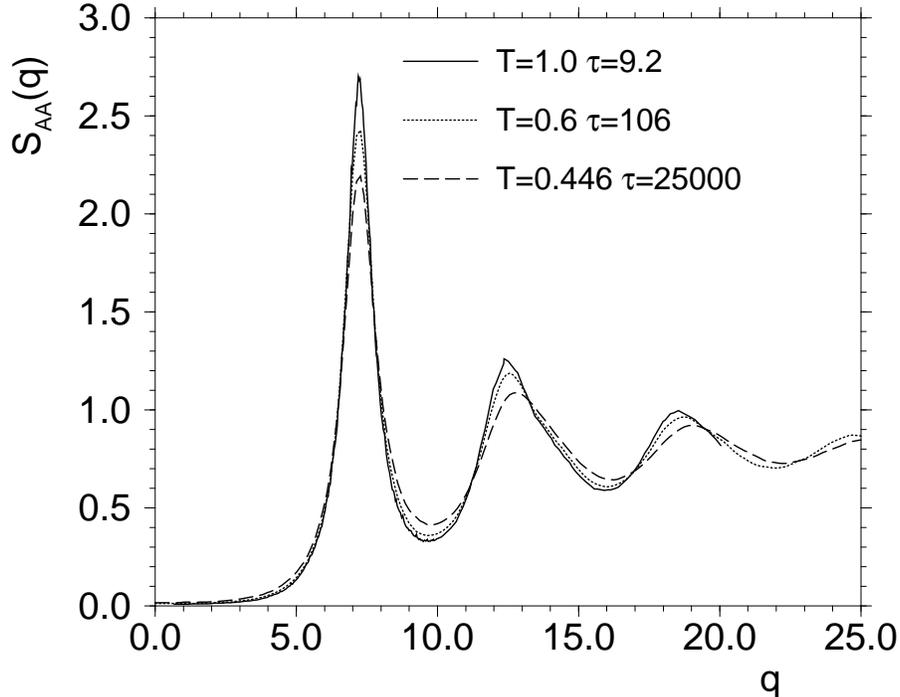}
\caption{Wave vector dependence of the structure factor for the A-A
correlation for different temperatures. Also included are the relaxation
times for the intermediate scattering function for the A-particles.}
\label{fig2}
\end{figure}

Before we study the {\it dynamical} properties of the system it is useful
to have a look at its {\it static} properties. In section~\ref{sec2} we
have mentioned that in the supercooled regime thermodynamic quantities
and structural quantities show only a weak temperature dependence. That
this is the case for the present system as well is demonstrated in
figure~\ref{fig2}, where we show the static structure factor\index{structure
factor} for the
A particles, $S_{\rm AA}(q)$, for different temperatures $T$. From
this figure we recognize that the $T-$dependence of $S_{\rm AA}(q)$ is
quite mild in that the main effect of a decreasing temperature is that
the various peaks become more pronounced and narrower. A similar weak
$T-$dependence is also found for the pressure and the total energy of the
system~\cite{kob95a}.  In order to demonstrate that, in the temperature
range shown, the {\it dynamics} of the systems changes strongly we have
included in the figure also the typical relaxation times, defined more
precisely below, at the different temperatures. From these numbers we
see that in this temperature range the relaxation dynamics slows down
by about two and a half orders of magnitude, a huge amount compared with
the weak temperature dependence of the structural quantity.

One of the simplest possibilities to study the dynamics of a liquid
is to investigate the time dependence of the mean-squared-displacement
(MSD)\index{mean squared displacement} which is defined by
\begin{equation}
\langle r^2(t) \rangle = \frac{1}{N_{\alpha}} \sum_{i=1}^{N_{\alpha}}
\langle |{\bf r}_i(t)-{\bf r}_i(0)|^2\rangle
\label{eq7}
\end{equation}
Note that here the sum over the particles of type $\alpha$ is not really
needed since in principle all particles of the same kind are statistically
equivalent.  However, in order to improve the statistics for the MSD
it is advisable to make the additional average over the particles of
the same kind.

\begin{figure}[hbt]
\vspace*{90mm}
\includegraphics{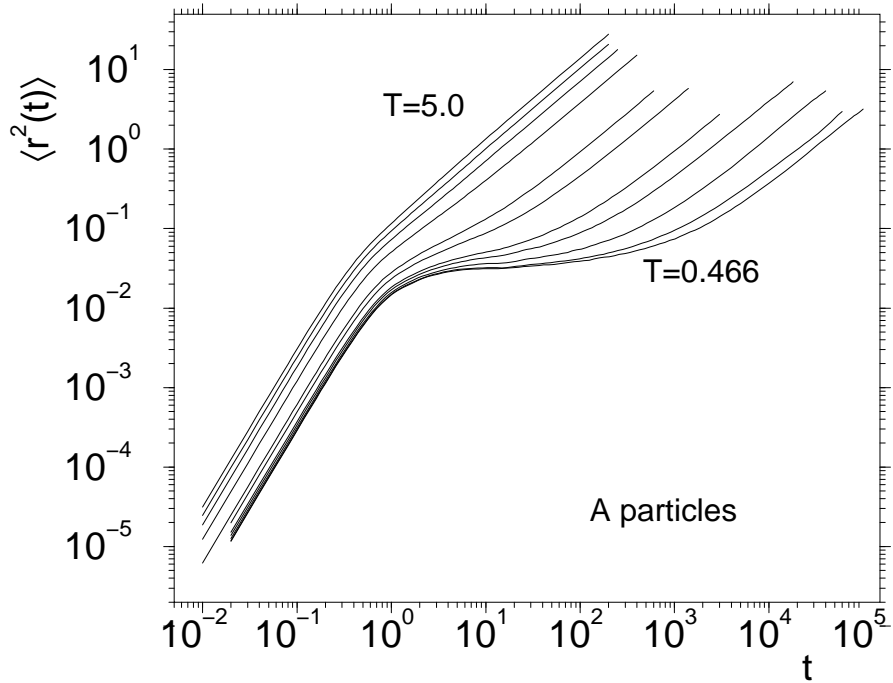}
\caption{Time dependence of the mean squared displacement for the A
particles for different temperatures.}
\label{fig3}
\end{figure}

In figure~\ref{fig3} we show the time dependence of the MSD for the
A particles at the different temperatures. Let us start our discussion
for the high temperatures, curves to the left. For very short times the
particle flies just ballistically\index{ballistic motion}, 
since on this time scale it does not
even realize that it is part of a many body system. Thus its position is
given by ${\bf r}_i(t)= {\bf r}_i(0)+{\bf v}_i(0)t$, where ${\bf v}_i(0)$
is its initial velocity. Thus the MSD is proportional to $t^2$, which
is the time dependence seen at short times (see figure).

After the ballistic flight the particle collides with its nearest
neighbors and thus its motion becomes diffusive\index{diffusion}, i.e. $\langle
r^2(t)\rangle=6Dt$, where $D$ is the diffusion constant. This diffusive
behavior is readily seen in the curves at long times. The two time regimes
just discussed are also found in the MSD for low temperatures. In addition
to them we see from the figure that a third regime is present in that
the ballistic and diffusive regime are separated by a time window in
which the MSD shows a plateau. This means that in this time regime the
particle does not significantly increase its distance from the point
it was at time zero. The physical picture behind this behavior is the
so-called ``cage effect''\index{cage effect}, i.e. the fact 
that on this time scale the
particle is trapped by its surrounding neighbors. Only at long times the
particle is able to escape this cage and to become diffusive again. Note
that the particles forming the cage are of course also caged since they
are surrounded by {\it their} neighbors. Hence it becomes clear that in
order to obtain a correct description of the dynamics of the particles
inside the cage and the breaking up of this cage, it is necessary to make
a {\it self-consistent} Ansatz for the motion of the particle and its
cage and MCT is one way to do this.

Since the intermediate scattering function $F(q,t)$\index{intermediate scattering
function} and its self
part $F_s(q,t)$ are of experimental relevance and are also the main
focus of MCT it is of course interesting to investigate their time and
temperature dependence. In figure~\ref{fig3} we show the time dependence
of $F_s(q,t)$ for different temperatures. The wave-vector $q$ is 7.25,
the location of the maximum in the static structure factor for the A-A
correlation. (For other wave-vectors the correlation functions look
qualitatively similar~\cite{kob95b}.) Also in this figure we find the different time
regimes that we have discussed in the context of the MSD.  At very short
times the correlator shows a quadratic time dependence, which corresponds
to the ballistic motion in the MSD. At high temperatures we see that
after this time regime $F_s(q,t)$ decays rapidly to zero, and it is found
that this decay is described well by an exponential. This behavior is
typical for a liquid at {\it high} temperatures and is not specific to
the present system. Also at low temperatures the quadratic time
dependence is found at short times. In contrast to the high temperature
case we find however at intermediate times a plateau, the origin of
which is again the cage effect\index{cage effect} that we have discussed before. Only
at very long times the correlation function decays to zero. This
ultimate decay is {\it not} given by an exponential, but by a so-called
Kohlrausch-Williams-Watts (KWW) law\index{Kohlrausch law} 
(also called stretched exponential),
i.e. by $A \exp(-(t/\tau)^{\beta})$, where the amplitude $A$, the
time scale $\tau$ and the Kohlrausch-exponent $\beta$ depends on the
wave-vector.

For the following discussion a bit of nomenclature is useful: The
time range in which the correlation function is {\it close} to the
mentioned plateau is called the $\beta-$relaxation regime\index{$\beta-$relaxation}. 
The time
window in which the correlator falls {\it below} the plateau is called
the $\alpha-$relaxation\index{$\alpha-$relaxation}. 
Note that the {\it late} $\beta-$relaxation
coincides with the {\it early} $\alpha-$relaxation regime.

\begin{figure}[hbt]
\vspace*{93mm}
\includegraphics{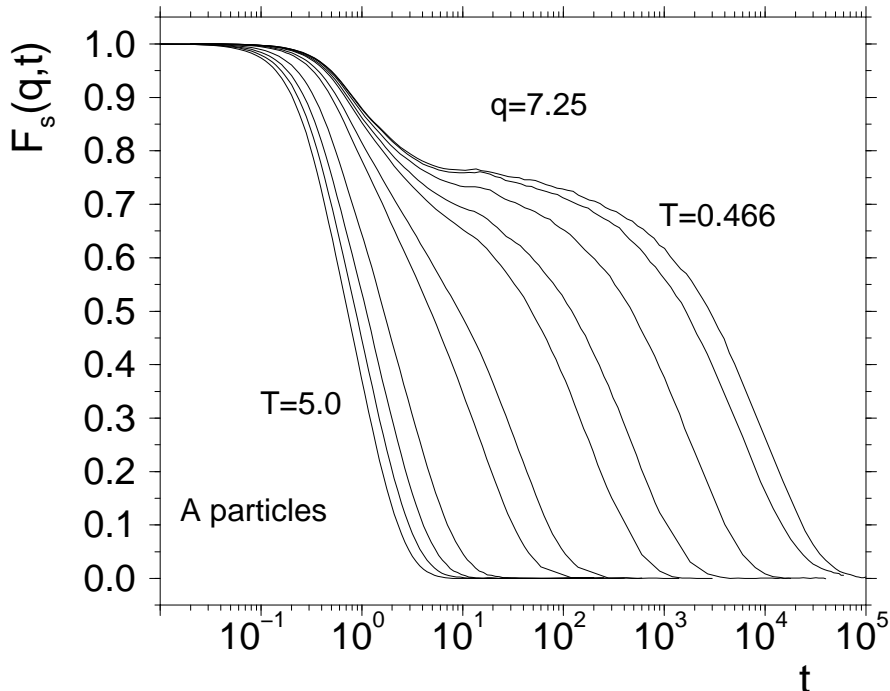}
\caption{Time dependence of the incoherent intermediate scattering
function of the A particles for all temperatures studied.}
\label{fig4}
\end{figure}

\begin{figure}[hbt]
\vspace*{90mm}
\includegraphics{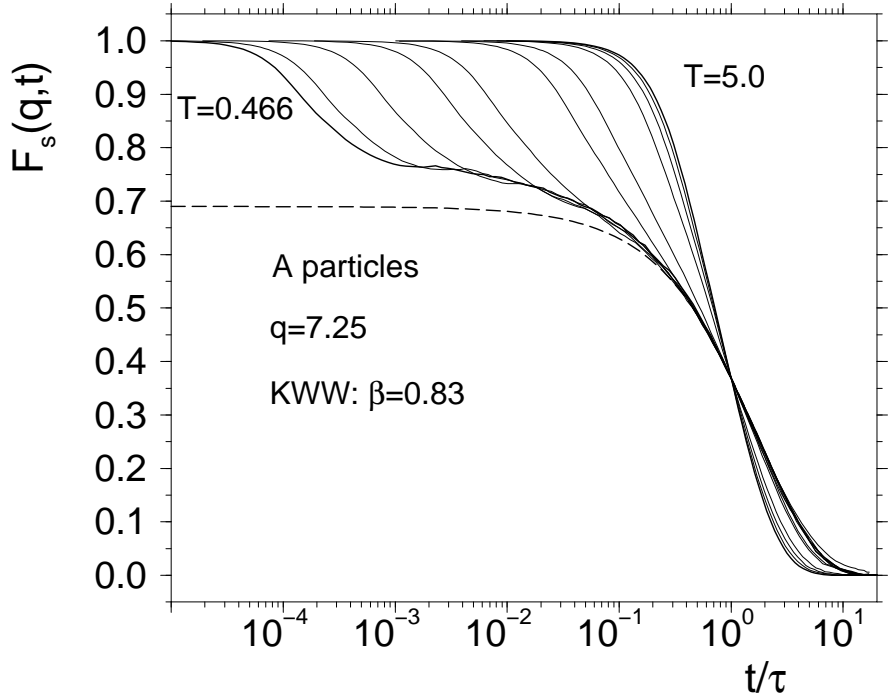}
\caption{The same correlation function as in figure~\protect\ref{fig4} versus
$t/\tau(T)$, where $\tau(T)$ is the $\alpha-$relaxation time of the system. The
dashed curve is a fit with the KWW function to the master curve in the
$\alpha-$relaxation regime.}
\label{fig5}
\end{figure}

MCT predicts that in the vicinity of the critical temperature $T_c$
the so-called time-temperature superposition principle \index{time-temperature
superposition principle} holds in the
$\alpha-$relaxation regime. This means that a time correlation function
$\phi(t)$ can be written as follows:
\begin{equation}
\phi(t,T)=\Phi(t/\tau(T)) \quad,
\label{eq8}
\end{equation}
where $\Phi$ is a master function which depends on $\phi$, and $\tau(T)$
is the $\alpha-$relaxation time at temperature $T$, which also depends on
$\phi$.  In order to check the validity of this prediction, we define the
$\alpha-$relaxation time as that time at which the correlator has fallen
to $1/e$ of its initial value. If equation~(\ref{eq8}) does indeed hold,
a plot of the correlation function versus the rescaled time $t/\tau(T)$
should give {\it in the $\alpha-$relaxation regime} a master curve. That
this is indeed the case is shown in figure~\ref{fig5}, where we show the
same data as in figure~\ref{fig4}, but this time versus $t/\tau(T)$. We
clearly see that the curves at low temperatures fall nicely onto a master
curve. In addition MCT predicts that the shape of this master curve can
be fitted well by the mentioned KWW law and a fit with this functional
form is included in the figure as well, showing that this law does indeed
fit our master curve very well.

\begin{figure}[hbt]
\vspace*{93mm}
\includegraphics{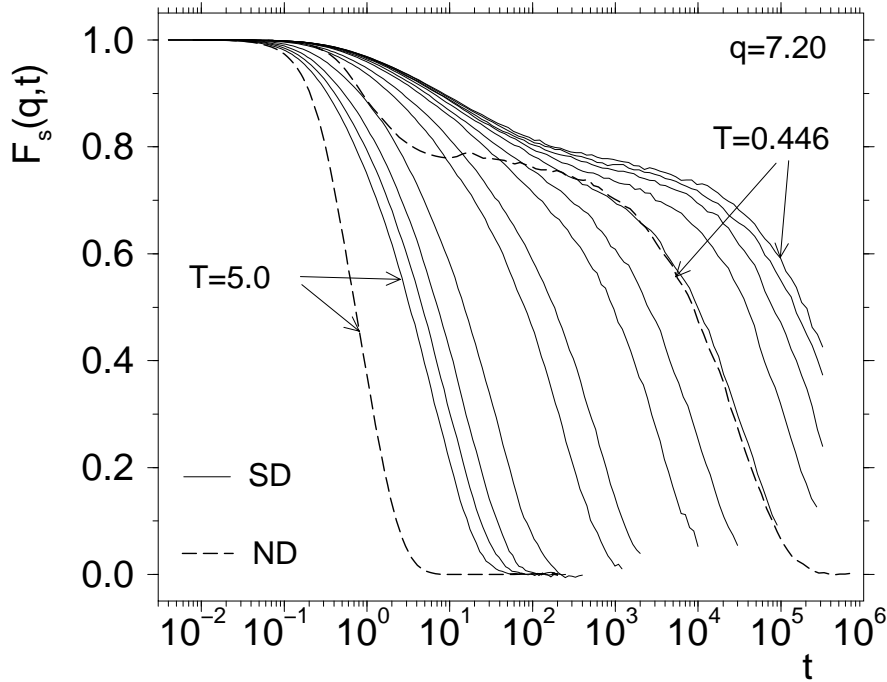}
\caption{The incoherent intermediate scattering function for the
stochastic dynamics for all temperatures investigated (solid lines). Bold
dashed line: $F_s(q,t)$ for the Newtonian dynamics at $T=5.0$ and $T=0.446$.}
\label{fig6}
\end{figure}

The results discussed so far are all for the ND, i.e. the dynamics in
which the microscopic motion of the particles is not damped. In order to see how
the relaxation dynamics changes if we have a strong damping we show
in figure~\ref{fig6} the self intermediate scattering function for the
stochastic dynamics (solid lines). The wave-vector is the same as the one
in figure~\ref{fig4}. We see that the time and temperature dependence
of the correlator is qualitatively the same as in the case of the ND.
However, a closer inspection shows important differences between the
two types of dynamics and in order to see them better we have included
in the figure also two curves for the ND (dashed lines). First of all
the time scale for the $\alpha-$relaxation is significantly larger for
the SD. Whereas at high temperature the SD relaxes slower by a factor
of about seven, this factor increases to a value around 30 at the lowest
temperature, and then stays constant~\cite{gleim98}.  Note however,
that apart from this change of time scale the $\alpha-$relaxation is
the same, in that the shape of the curves as well as the height of the
plateau for the ND and SD is the same (see also figure~\ref{fig6}).
This is exactly what is expected within MCT in that the theory  predicts
that at temperatures around $T_c$ the temperature dependence of the
dynamics is independent of the microscopic dynamics, {\it apart from a
system universal constant factor}.

\begin{figure}[hbt]
\vspace*{93mm}
\includegraphics{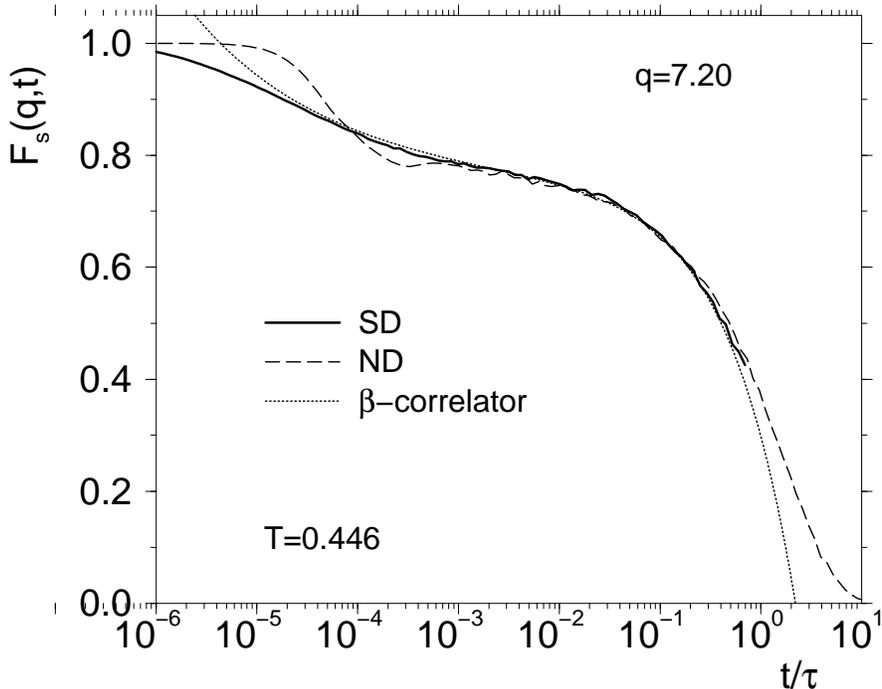}
\caption{The incoherent intermediate scattering function for the ND and SD, dashed
and solid lines, respectively, versus the scaled time $t/\tau(T)$. $T=0.446$. The
dotted curve is a fit with the $\beta-$correlator from MCT.}
\label{fig7}
\end{figure}

Although the relaxation of the curves away from the plateau is independent
of the microscopic dynamics the approach of the curves {\it to} the
plateau depends on it. In particular we see that for the ND this approach
is very abrupt whereas it is very gentle for the SD. The reason for this
difference is that in the SD the phonon-like motion of the particles
is strongly damped and thus the particles explore their cage in a much
gentler way as they do in the ND. In order to investigate this part of the
$\beta-$relaxation dynamics in more detail we show in figure~\ref{fig7}
the SD curves from figure~\ref{fig6} and the ND curve at low temperature
versus $t/\tau(T)$. From this figure we see that the curves for the two
different kinds of dynamics do indeed fall on top of each other in the
$\alpha-$relaxation regime but that they show the mentioned differences
in the early $\beta-$relaxation regime.

MCT\index{mode-coupling theory} 
predicts that {\it in the $\beta-$relaxation regime}\index{$\beta-$relaxation} 
the shape
of the master curve is not arbitrary, but is given by the so-called
$\beta-$correlator\index{$\beta-$correlator}, a function which is the solution of a certain
integral equation~\cite{gotze99,mct_reviews}. This integral equation,
and hence its solution, depend on one parameter $\lambda$, the
so-called ``exponent parameter''\index{exponent parameter}. 
The value of $\lambda$ can be calculated
from the structure factor and has for the present system the value
0.708~\cite{nauroth97}. Using this value of $\lambda$ it is possible
to solve the mentioned integral equation and thus to calculate the
$\beta-$correlator. In figure~\ref{fig7} we have included (bold dotted
line) the best fit with this $\beta-$correlator and we recognize that
this functional form gives a very good description of the correlators in
the vicinity of the plateau.  In particular we see that in the case of
the SD the fit is also good in the early $\beta-$regime, thus showing
that the damping of the motion leads to a much better agreement with
the theory. The reason for this is that if no damping is present, the
dynamics at short times, which is governed by phonon-like motion, strongly
interferes with the relaxation in the early $\beta-$relaxation regime and
thus leads to the observed discrepancy between the $\beta-$correlator
and the curve from the ND. However, if a one takes into account in the
theory this phonon-like dynamics a good agreement between the theory
and the ND curves is found also~\cite{nauroth99}. Thus we can conclude
that MCT is able to give correct description of the $\beta-$relaxation
dynamics on a qualitative as well as quantitative level.

\begin{figure}[hbt]
\vspace*{97mm}
\includegraphics{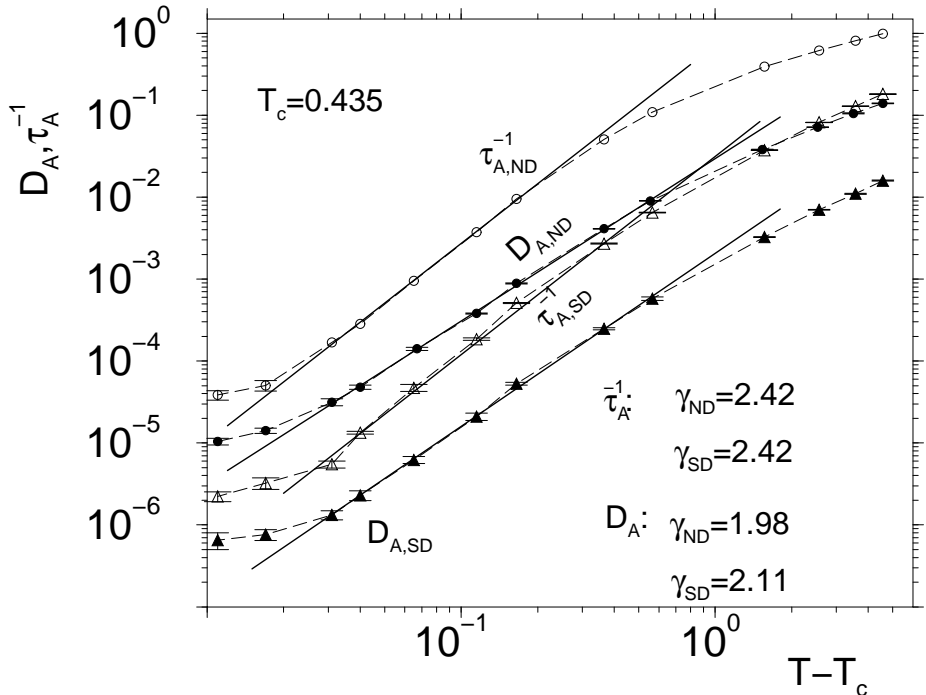}
\caption{Diffusion constant and inverse of the $\alpha-$relaxation time
versus $T-T_c$. Open and filled symbols correspond to the Newtonian and
stochastic dynamics, respectively. The bold straight lines are fits to
the data with power-laws.}
\label{fig8}
\end{figure}

We now turn our attention to the temperature dependence of the diffusion
constant and the relaxation times. MCT predicts that in the vicinity of
$T_c$ these quantities should show a power-law dependence, i.e.
\begin{equation}
D(T) \propto \tau^{-1}(T) \propto (T-t_c)^{-\gamma}
\label{eq9}
\end{equation}
where the exponent $\gamma$ can be calculated from the exponent parameter
$\lambda$ and is found our system to be 2.34~\cite{nauroth97}. In
figure~\ref{fig8} we show the temperature dependence of the diffusion
constant and the $\alpha-$relaxation time $\tau$ for the A particles
for the case of the ND and SD. In order to check for the presence of
the power-law given by equation~(\ref{eq9}) we plot these quantities
versus $T-T_c$, where the critical temperature $T_c$ was used as a
fit parameter~\footnote{We mention that in principle it is possible to
calculate the value of $T_c$ within MCT. However, it has been found that
the theoretical value, $T_c=0.92$, is very far from the one determined
from the correlation functions ($T_c=0.435$)~\cite{nauroth97}. This
discrepancy is not a particularity of the present system but reflects
the fact that MCT seems to have difficulty to estimate this quantity
with high accuracy.} We see that in the supercooled regime the data
can be fitted very well by such a power-law. In particular we find
that the exponent $\gamma$ of the power-law for the relaxation time is
independent of the microscopic dynamics, see the values for $gamma$
in the the figure, and the same is true also for the exponents for
the diffusion constants. However, in contrast to the prediction of the
theory (see equation~\ref{eq8}), the exponent for the relaxation time
is {\it not} the same as the ones for the diffusion constant\index{diffusion
constant}. The
reason for this is likely the fact that the system is dynamically
heterogeneous~\cite{dyn_het_exp,dyn_het_sim}\index{dynamical heterogeneities}, 
i.e. it has regions in which
the dynamics of the particles is significantly faster than in other
regions. Since within MCT it is not possible to take into account such
dynamical differences, due to the mean-field like nature of the theory,
the prediction of MCT for the temperature dependence of the product $D(T)
\tau(T)$ is, {\it for the present system}, not correct.

\begin{figure}[hbt]
\vspace*{90mm}
\includegraphics{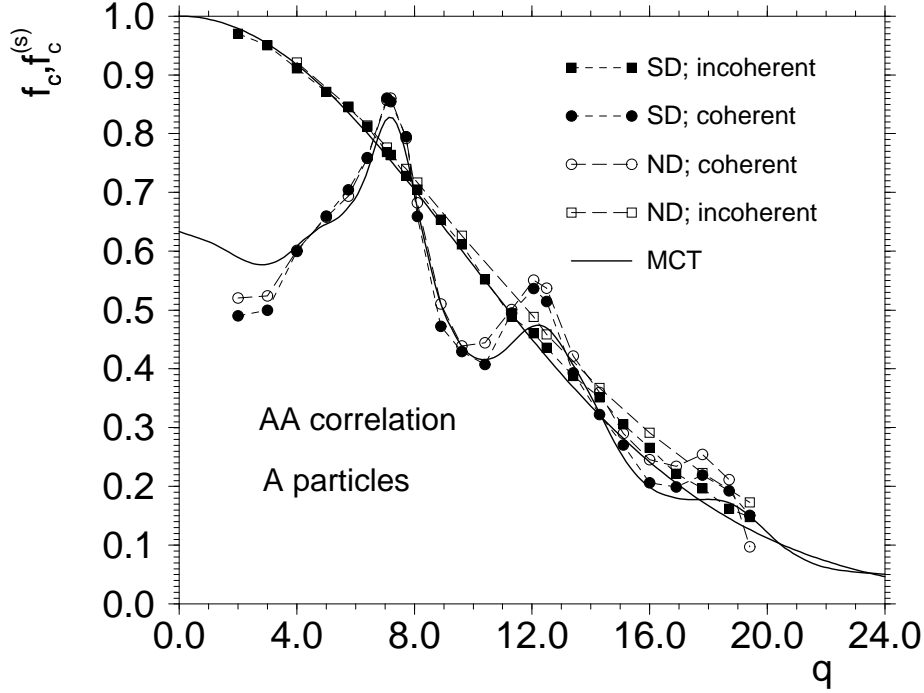}
\caption{Wave-vector dependence of the nonergodicity parameter for the
coherent and incoherent intermediate scattering function (circles and
squares, respectively). The open and closed symbols correspond to the
Newtonian and stochastic dynamics, respectively.}
\label{fig9}
\end{figure}

So far we have only tested the applicability of MCT on a {\it
qualitative} level.  These types of checks, and many more, can be
done for all systems for which the dynamics has been studied in a
temperature range in which the time scale of the dynamics changes
considerably and in Ref.~\cite{gotze99} many of these tests are
discussed. For simple liquids also {\it quantitative} tests are
possible if the static structure factor is known with sufficiently
high accuracy (e.g. 1\% accuracy for wave-vectors between $0.1 q_0
\leq q \leq 3 q_0$, where $q_0$ is the location of the maximum
in $S(q)$). For this one has to solve the wave-vector dependent
mode-coupling equations~(equations~(\ref{eq3})-(\ref{eq5})), using the
static structure factor as input. This has been done for hard sphere
system and the theoretical results compare nicely with the ones from
experiments on colloidal particles~\cite{gotze99,colloid_tests}. Similar
calculation have also been done for soft sphere systems~\cite{barrat90}
and water~\cite{fabbian99}. Here we will discuss the results for the
present Lennard-Jones mixture. One quantity which is relatively simple
to calculate is the value of the so-called critical nonergodicity
parameter\index{nonergodicity parameter}, which is the height of the
plateau in a time correlation function {\it at $T_c$}. Note that for
the case that the correlation function is the intermediate scattering
function this quantity will depend on the wave-vector as well as on the
type of particle. In figure~\ref{fig9} we show the $q$ dependence of the
nonergodicity parameter for the case of $F_s(q,t)$ for the A particles as
well as for $F(q,t)$ for the A-A correlation (ND, open symbols). We see
that the coherent part shows an oscillatory behavior which is in phase
with the structure factor. The reason for this is that the structure
of the liquid is very stiff on the length scale of the interparticle
distances thus leading to a high plateau in the time correlation function,
i.e. a large nonergodicity parameter.

In order to check whether the value of the nonergodicity parameters
depend on the microscopic dynamics we have included in the figure also
the data for the SD. We see that the curves for the SD are very close to
the ones for the ND and thus conclude that the height of the plateau is
independent of the microscopic dynamics, in agreement with the prediction
of MCT. Also included in the figure is the theoretical prediction from
MCT for the nonergodicity parameter (solid lines)~\cite{nauroth97}. We
see that these theoretical curves fall nicely onto the data points from
the simulations thus demonstrating that the theory is indeed able to
make also correct quantitative predictions. It should be appreciated
that no free fit parameter of any kind was used to calculate the
theoretical curves. We also mention that a similar good agreement between
simulation and theory is obtained for the nonergodicity parameters of
the intermediate scattering function for the A-B and B-B correlation as
well as for the B particles.

\begin{figure}[hbt]
\vspace*{90mm}
\includegraphics{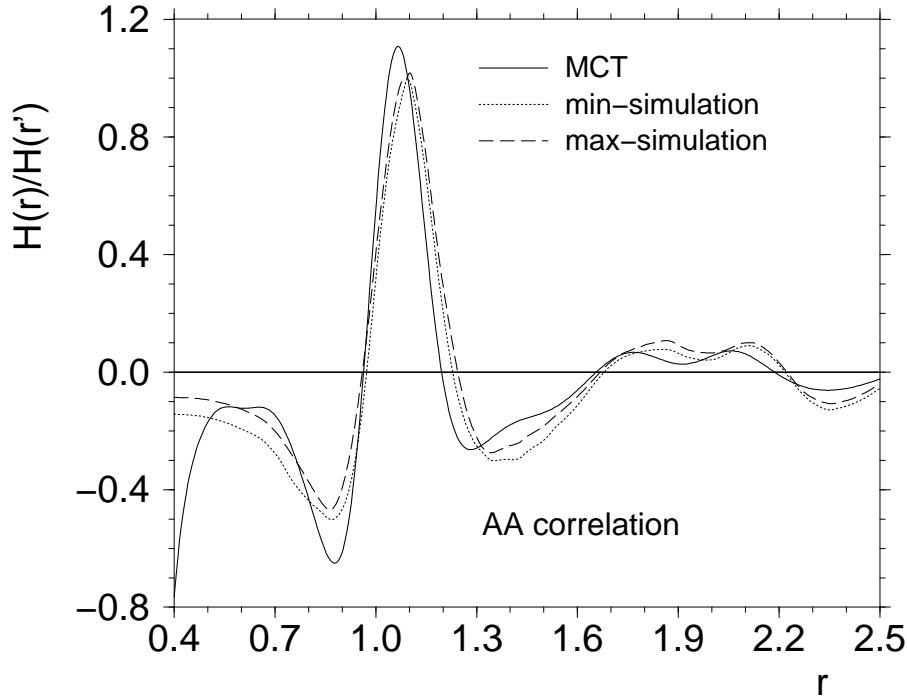}
\caption{$r-$dependence of the critical amplitude for $G_d^{\rm AA}(r,t)$
for the A-A correlation as determined from the simulation and the
mode-coupling theory.}
\label{fig10}
\end{figure}

As a further test of a quantitative prediction of the theory we will
finally discuss some results of the dynamics in the $\beta-$relaxation
regime\index{$\beta-$relaxation}. 
MCT predicts that as long the time correlation function is close
to the plateau it can be written as follows:
\begin{equation}
\phi_l(t)=f_l^c+ h_l G(t) \quad ,
\label{eq10}
\end{equation}
where $l$ is just an index labeling the correlator, $f_l^c$ is the
nonergodicity parameter discussed above, $h_l$ is the so-called
critical amplitude\index{critical amplitude}, 
and $G(t)$ is a system universal function, i.e. it
is independent of $l$. The physical contents of this equation is that
{\it in the $\beta-$regime} all time correlation function have the
same time dependence, namely the one given by the function $G(t)$.
Therefore equation~(\ref{eq10}) is also sometimes called the
``factorization property''. In order to check the validity of this
prediction we have used for the functions $\phi_l(t)$ the distinct part
of the van Hove correlation functions, $G_d^{\alpha\beta}(r,t)$. (Note
that therefore the space variable $r$ takes to role of the index $l$
in equation~(\ref{eq10}).) These space-time correlations are defined by
\begin{equation}
G_{d}^{\alpha\alpha}(r,t)=
n g_{\alpha \alpha}(r,t) =
\frac{N_{A}+N_{B}}{N_{\alpha}(N_{\alpha}-1)}
\left\langle\sum_{i=1}^{N_{\alpha}} \sum_{j=1}^{N_{\alpha}}\mbox{ }\!\!'\,
\delta \left(r-|\mbox{\boldmath
$r$}_{i}(0)-\mbox{\boldmath $r$}_{j}(t)|\right) \right\rangle
\label{eq11}
\end{equation}
and
\begin{equation}
G_{d}^{AB}(r,t)=
n g_{AB}(r,t) =
\frac{N_{A}+N_{B}}{N_{A}N_{B}}
\left\langle\sum_{i=1}^{N_{A}} \sum_{j=1}^{N_{B}}
\delta \left(r-|\mbox{\boldmath
$r$}_{i}(0)-\mbox{\boldmath $r$}_{j}(t)|\right) \right\rangle
\quad ,
\label{eq12}
\end{equation}
where $n$ is the particle density of the system. Note that for $t=0$
these functions are just the usual (partial) radial distribution functions
and hence $G_d^{\alpha\beta}(r,t)$ can be considered as a generalization
of the latter to the time domain. In Ref.~\cite{kob95a} we have shown
that for this set of corelation functions the factorization property is indeed
fulfilled, i.e. that in the $\beta-$relaxation regime the correlators have the
form given by equation~(\ref{eq10}). From that equation it follows immediately
that the following equation holds for all values of $r$:
\begin{equation}
\frac{G_d^{\alpha\beta}(r,t)-G_d^{\alpha\beta}(r,t')}
{G_d^{\alpha\beta}(r',t)-G_d^{\alpha\beta}(r',t')} = \frac{H^{\alpha\beta}(r)}
{H^{\alpha\beta}(r')} \quad ,
\label{eq13}
\end{equation}
where $H^{\alpha\beta}(r)$ is the critical amplitude for the function
$G_d^{\alpha\beta}(r,t)$, and $r'$ is arbitrary, and $t$ and $t'$ are
arbitrary times in the $\beta-$regime. Since the factorization property
holds it thus becomes possible to determine from the simulation the $r$
dependence of the ratio $H^{\alpha\beta}(r)/H^{\alpha\beta}(r')$. In
figure~\ref{fig10} we show an upper and lower bound for this function
(for the case of the A-A correlation), as it was determined from the
simulation and we see that this is a nontrivial function of $r$. Also
included in the figure is the theoretical value for this ratio and we
see that this curve reproduces well the one from the simulation (also
in this case no free fit parameter exists). Thus this is more evidence
that MCT is not only able to make correct qualitative predictions but
also quantitative ones.

Many more tests have been done in order to find out to what extend MCT
is able to predict the dynamics of this system at low temperatures. The
outcome  of these tests is that the theory is indeed able to give a good
description of this dynamics. Since a similar conclusion has been reached
for the case of hard spheres, where the theoretical predictions have been
compared with experiments on colloidal systems, we thus can conclude that
MCT is able to describe the dynamics of simple liquids on a qualitative
as well as quantitative level. To what extend this is the case also for
more complex systems, such as molecules with odd shapes or systems with
long range interactions, is currently still a matter of investigation. The
results for water~\cite{sciortino_water}, a triangular shaped molecule,
and silica~\cite{horbach}, a system with long range interactions, look,
however, promising.

\section{Out of Equilibrium Dynamics}\index{out of equilibrium dynamics}

The results discussed in the previous section are concerned with the
dynamics of the supercooled liquid {\it in equilibrium}. We have seen
that with decreasing temperature this dynamics slows down and hence it is
clear that there will exist a finite temperature below which the system
cannot be equilibrated anymore within the time scale of the experiment or
the computer simulation. Hence the system will fall out of equilibrium,
i.e. undergo a glass transition\index{glass transition}. As this is a 
purely kinetic phenomenon,
the temperature at which this happens is not intrinsic to the system
such as, e.g. its melting point, and thus can be changed by choosing a
different experimental time scale. For the sake of convenience we will
call this temperature the glass temperature $T_g$, despite the fact that
we have defined this term in section~\ref{sec2} differently. Since below $T_g$ the
system is no longer able to relax one might expect that the motion of the
particles essentially stops, apart from their vibration inside the cage,
i.e. that relaxation no longer takes place.  In order to check whether
this expectation is born out we investigate in this section the dynamics
of a system after a quench below $T_g$. As we will see, even below $T_g$
relaxation takes place but its nature is very different from the one
in equilibrium, i.e. above $T_g$. In particular it is found that the
properties of the system, such as its structure or relaxation times,
change with time. Therefore it is customary to say that the system
is aging\index{aging}.

Although experiments on aging materials have been done since many years,
mainly on polymeric systems, since they often show very pronounced
aging effects such as the material becoming more brittle with
time~\cite{aging_exp}, their theoretical description was done only on
a phenomenological level. Only in recent times strong efforts have been
undertaken in order to understand this situation within a well defined
theoretical framework~\cite{cugliandolo93,aging_theo}. However, from a
theoretical point of view these aging systems are still understood in
much less detail than it is the case for the (supercooled) equilibrium
system and very often only predictions of very general nature can be made.

In the following we will discuss some results of simulations which have
been done in order to investigate the dynamics of a simple glass-former
which has been quenched below $T_g$. The system of interest is the same
binary Lennard-Jones \index{Lennard-Jones} mixture whose equilibrium
properties we have discussed in the previous section. There we have seen
that for this system the relaxation times close to the MCT temperature
$T_c$ start to become comparable with the longest runs of present days
simulations. Thus from a practical point of view the glass transition
(on the computer!)  takes place around $T_c (=0.435)$. To investigation
the dynamics of the system below $T_c$ we equilibrated it at a temperature
$T_i>T_c$ and then quenched it at time zero to a final temperature $T_f
\leq T_c$. This quench was done by coupling the system every 50 time
steps to a stochastic heat bath, i.e. all the velocities of the particles
are substituted with ones drawn from a Maxwell-Boltzmann distribution
corresponding to a temperature $T_f$.

When one investigates the properties of an aging system it is useful to
distinguish between two types of observables: the so-called ``one-time
quantities'' and the ``two-times quantities''. The former term refers to
observable which {\it in equilibrium} are constants, such as the density
(in a constant pressure experiment), the total energy of the system,
or the structure (as measured, e.g., by the structure factor). In the
out of equilibrium situation the value of such observables depend on the
time since the quench and hence they depend on {\it one time}. Two-times
quantities are time correlation functions which in equilibrium depend on a
time difference, such as the mean squared displacement or the intermediate
scattering function. Since in nonequilibrium the time elapsed since the
quench has to be taken into account also, such quantities will depend
on {\it two} times in the aging system.

In agreement with theoretical predictions it has been found,
see e.g.~\cite{andrejew96,kob97,kob99a}, that most one ``one-time
quantities'' depend only weakly on time. Examples investigated were
the total energy of the system, the radial distribution function
or the pressure~\footnote{We mention, however, that certain one-time
quantities can show a sufficiently strong time dependence so that they can
be used to characterize the aging system very well. Examples of such
observables are discussed in Ref.~\cite{kob99b}}. In contrast to this
the two-times quantities showed a very strong time dependence (see also
Ref.~\cite{parisi97} for similar results for a soft sphere system.)
A typical
example of a correlator that has such such a strong time dependence is
$C_k(t_w+\tau,t_w)$, the generalization of the incoherent intermediate
scattering function\index{intermediate scattering function}, 
see equation~(\ref{eq6}), to the out of equilibrium
situation. Thus $C_k(t_w+\tau,t_w)$ is given by
\begin{equation}
C_k(t_w+\tau,t_w)= \frac{1}{N}\left\langle \sum_j^N \exp(i{\bf q}\cdot ({\bf
r}_j(t_w+\tau)-{\bf r}_j(t_w)) \right\rangle
\quad,
\label{eq14}
\end{equation}
where $t_w$ is the time between the quench and the start of the
measurement and hence is also called ``waiting time''\index{waiting
time}.  Thus the meaning of this time correlation function is that a
density fluctuation which is present at a time $t_w$ after the quench
is correlated with a density fluctuation at a time $\tau$ later.

\begin{figure}[hbt]
\vspace*{90mm}
\includegraphics{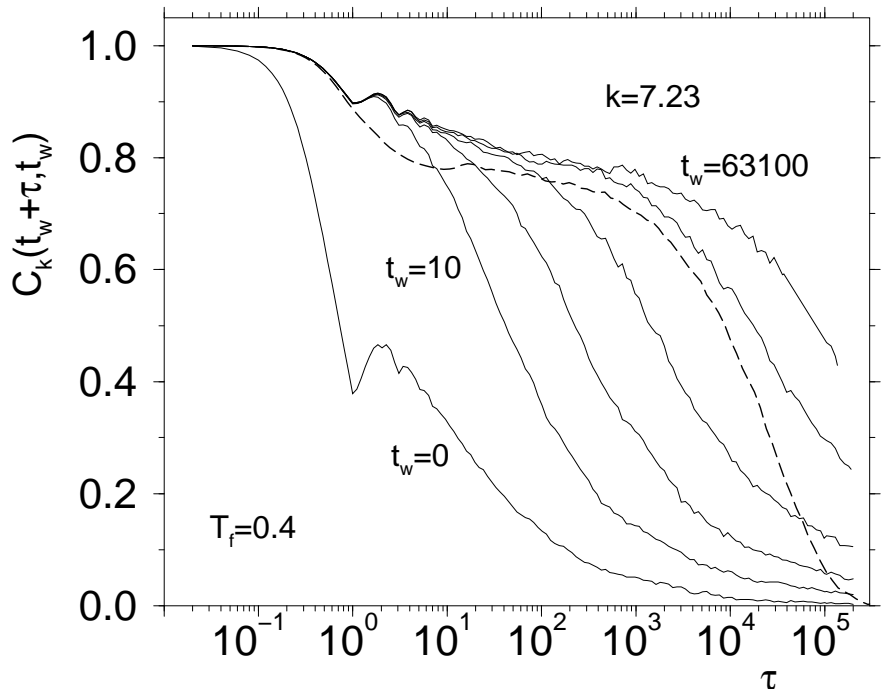}
\caption{Time dependence of the generalization of the incoherent
intermediate scattering function to the out-of equilibrium situation
for waiting times $t_w=0,$ 10, 100, 1000, 10000, and 63100. $T_f=0.4$.
The dashed line is the equilibrium correlation function at $T=0.446$.}
\label{fig11}
\end{figure}

In figure~\ref{fig11} we show the $\tau$ dependence of $C_k$ for different
waiting times and the A particles. The wave-vector is $k=7.23$, i.e. the
location of the maximum in the static structure factor for the A-A
correlation. For small values of $T_w$ the curves rapidly decay to zero.
With increasing $t_w$ the curves show at intermediate times a plateau
and go to zero only at long times. We see that if $t_w$ is not too small
the approach of the curves {\it to} the plateau is independent of $t_w$,
whereas the time at which they start to fall below the plateau depends
on the waiting time. In Ref.~\cite{kob97} we have shown that the time at
which the curves leave the master curve is approximately proportional to
$t_w^{\alpha}$, with $\alpha=0.9$. Thus we find that $C_k$ does indeed
show a strong waiting time dependence, as it is theoretically expected for
a two-time quantity.  \footnote{Note that the oscillation at $t=1$ and
multiples of it originate in the coupling of the system to the external
heat bath and thus are not really an intrinsic feature of the aging system.}

In view of the fact that we are at a very low temperature it might
be a bit surprising to see that all the curves approach zero at long
times since from the relaxation behavior at {\it equilibrium} one would
expect that within the time span shown the curves should just fall on the
plateau and then stay in its vicinity (see figure~\ref{fig4}). However,
one should recall what is happening during the quench: At time zero
the configuration of the particles corresponds to one which is typical
for the high temperature $T_i$. Due to the quench the system now tries
to equilibrate and to do this it has to move to a part in configuration
space which is typical for configurations at $T_f$. It is this motion of
the system in configuration space which leads to the rapid decay of the
correlation function. If the waiting time since the quench is large, the
system is able to find configurations which are already closer
to the ones typical for $T_f$ and thus the driving force for further
exploration decreases. Hence the (out of equilibrium) relaxation becomes
slower and slower and thus it takes the correlation functions more and
more time to decay to zero.

Also included in the figure is the {\it equilibrium} curve at $T=0.446$
(bold dashed line). Although the shape of this curve is {\it qualitatively}
similar to the aging curves for long waiting times, a closer inspection
shows that there are important differences. For example the approach
of the curves to the plateau is much more rapid in the equilibrium
case than in the nonequilibrium case. Also at long times significant
differences are found. In figure~\ref{fig5} we have shown that at
long times the equilibrium curve can be fitted well with a KWW
law\index{Kohlrausch law}.  This is not the case for the out of
equilibrium case where it is found that the correlators show a power-law
dependence on time with an exponent which decreases with decreasing
wave-vector~\cite{kob99a}, which is, however, independent of the
waiting time.

\begin{figure}[tbh]
\vspace*{90mm}
\includegraphics{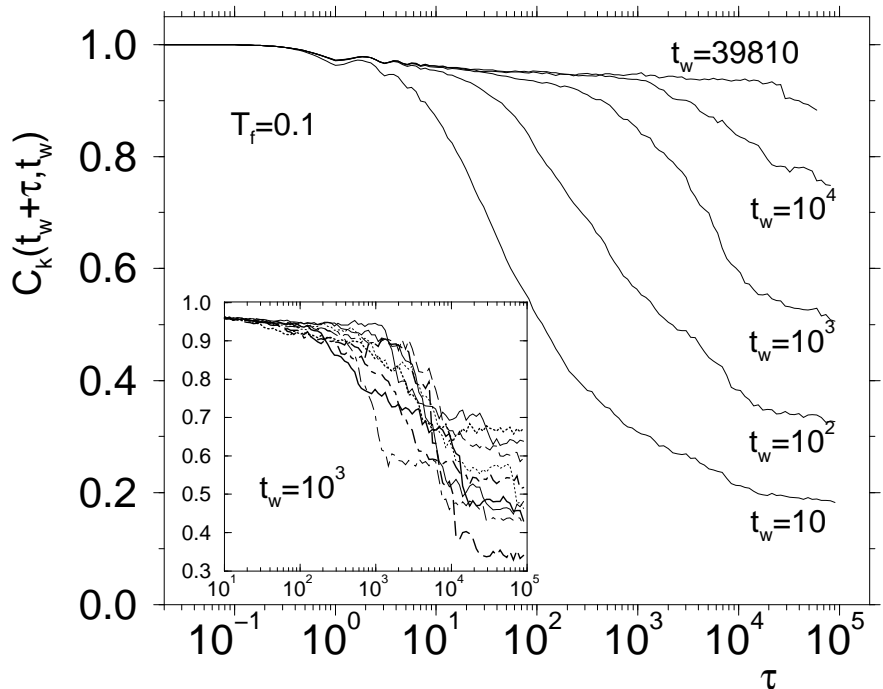}
\caption{Time dependence of $C_k(t_w+\tau,t_w)$ for different 
waiting times. $T_f=0.1$}
\label{fig12}
\end{figure}

The results discussed so far are for a quench to $T_f=0.4$, i.e. a
temperature which is quite close to the critical temperature of MCT. If
the final temperature is significantly lower, the relaxation behavior
can be quite different from the one with higher $T_f$. This is shown
in figure~\ref{fig12} where we show the same correlation function as in
figure~\ref{fig11}, but this time for $T_f=0.1$. From this figure we see
that, for long waiting times, the correlators at {\it short} times look
qualitatively similar to the ones for $T_f=0.4$. The main difference is
that the height of the plateau is higher, which is reasonable since this
height is, even in the out of equilibrium situation, related to the size
of the cage that each particle feels, and it can be expected that this
size is proportional to $1-T_f$.

For long times $\tau$ the curves for high and low values of $T_f$ are
different also on a qualitative level, in that the ones for $T_f=0.1$
show a {\it second} plateau. In the inset of figure~\ref{fig12} we show
the curves for $t_w=10^3$ for the {\it individual} runs. We now recognize that
most of these curves show at a time between $10^2-10^4$ time units one or
more sharp drops which are then followed by a regime in which the curves
is almost constant. It is this constant part which gives rise to the
second plateau in the average curve shown in the main figure whereas the
sharp drops average out to a much less sharp decrease in the mean curve.

In order to find out about the microscopic reason for the sharp drops
and the subsequent plateaus we have compared the configurations just
before the drop with the ones just after the drop~\cite{kob99a}. We have
found that the fast relaxation is due to the fact that around 10\% of the
particles, i.e. $O(100)$, undergo a sudden, quite cooperative motion in
which the particles move by around 0.2-0.5 of their diameter. Despite
the smallness of this motion, its cooperative nature leads to the
observed fast drop in the correlation function. The likely reason for
the occurrence of this cooperative motion is that, due to the quench,
the system has build up an internal stress and it seems that the most
efficient way to release this stress \index{stress release}is to rearrange
the particles in a cooperative way. Thus the situation is similar to an
earthquake where stress is released in a similar way.

Finally we discuss some very interesting results concerning the connection
between the time correlation functions and the response of the system to
an external perturbation. In equilibrium this connection is given by the
fluctuation dissipation theorem (FDT) \index{fluctuation dissipation theorem} 
which says the following: Consider
an observable $A$ and the associated normalized time auto-correlation
function $C(t)=\langle A(t)A(0)\rangle/\langle A(0) A(0)\rangle$. If
the system is perturbed with a field conjugate to the observable $A$
the response function $R(t)$ \index{response function} 
is given by $R(t)= -{1\over k_B T} {dC\over dt}$,
where $T$ is the temperature of the system. Thus in equilibrium the FDT
relates the time derivative of the correlation function with the response
and the factor is the inverse temperature.

In the derivation of the FDT it is required that the system is time
translation invariant, an assumption which is clearly not fulfilled
in the out of equilibrium situation. Hence the FDT does not hold
anymore and it has been proposed that the FDT should be generalized as
follows~\cite{cugliandolo93}: Since the correlator depends on two times,
also the response will depend on two times. Thus we have, assuming
$t'\geq t$,
\begin{equation}
R(t',t) = {1\over k_B T} X(t',t) {\partial C(t',t) \over \partial t}
\quad ,
\label{eq15}
\end{equation}
where the function $X(t,t')$ is defined by this equation. In the context
of mean-field spin glasses it has been shown that in the limit $t_w,\tau
\rightarrow \infty$, $X(t_w+\tau,t_w)$ is a function of the correlation
function $C$ only, i.e.
\begin{equation}
 X(t_w+\tau,t_w)=x(C(t_w+\tau,t_w)),
\label{eq16}
\end{equation}
where the function $x$ is now a function of {\it one} variable only.
(Here $t_w$ is again the time since the quench.) Within mean-field it is
expected that the function $x(C)$ is equal to $-1.0$ if $C$ is larger than
the plateau value, i.e. that for these short times the FDT holds.  For
times such that $C$ has fallen below the plateau it is expected that $x$
is larger than $-1.0$, i.e. the FDT is ``violated''. (The quotes reflect the
fact that of course the FDT is not violated, since it is not supposed to
hold.) The reason for the interest in the function $X(t',t)$ is twofold:
First, we see from equation~(\ref{eq15}) that $-k_BT/X$ is something
like an effective temperature\index{effective temperature}. 
Thus, if the time and temperature dependence of $X$
is known it might become possible to use thermodynamics concepts also for
the out-of-equilibrium system. Secondly, in the context of spin glasses
it has been found that the dependence of $x(C)$ can be used to classify
various types of spin glasses (see, e.g., reference~\cite{ricci99} for a
nice discussion on this). Thus by measuring $x(C)$ for a structural glass,
it might become possible to connect the properties of a {\it structural}
glass, such as the present Lennard-Jones system, with a {\it spin} glass.

Since the correlation function of interest is the generalization of the
incoherent intermediate scattering function, i.e. the correlation of a
density fluctuation, we need a method to measure the response function to
such fluctuations. Theoretically one could apply an external field with
wave-vector ${\bf q}$ which couples to the position of one particle and
see how this perturbation affects the density distribution. However,
this approach would lead to a very poor statistics and thus a more
efficient method has to be used, the details of which is described in
references~\cite{kob99a,barrat99}. That procedure allows one to measure the
integrated response $M(t_w+\tau,t_w)$ with reasonable accuracy, where
$M(t_w+\tau,t_w)$ is given by:
\begin{equation} 
M(t_w+\tau,t_w)=   \int_{t_w}^{t_w+\tau} R(t_w+\tau, t) dt \quad .
\label{eq17}
\end{equation}
Using equations~(\ref{eq15}) and (\ref{eq16}) one can rewrite this as
\begin{equation}
M(t,t') = M(C) = \frac{1}{k_BT}\int_C^1 x(c) dc\quad.
\label{eq19}
\end{equation}
From this equation it becomes clear that a parametric plot of the
integrated response versus the correlator will give us the information
about the integrant $x(c)$ and hence the factor $X(t_w+\tau,t_w)$.

\begin{figure}[hbt]
\vspace*{99mm}
\includegraphics{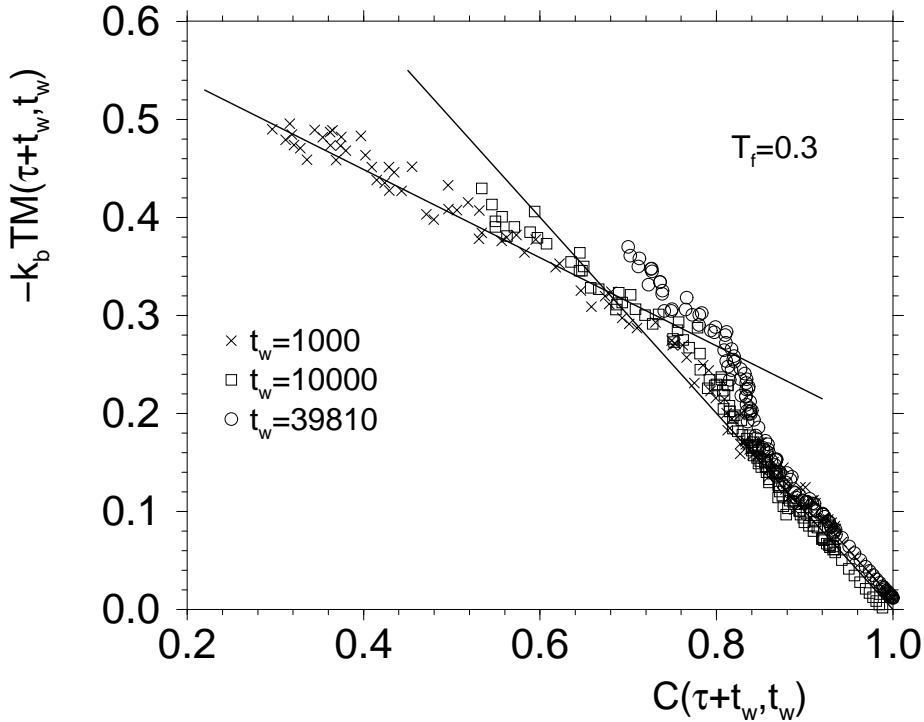}
\caption{Parametric plot of the integrated response $M_k(t_w+\tau,\tau)$
versus the correlation function $C_k(t_w+\tau,\tau)$ for different
waiting times and $T_f=0.3$. The straight lines have slopes around $-1$ and
$-0.62$.}
\label{fig13}
\end{figure}

In figure~\ref{fig13} we show such a parametric plot for different
waiting times and from it we can recognize the following things: For
short times, i.e. those points at which $C(t_w+\tau,t_w)$ is large, the
data points are compatible with a straight line with slope $-1$, i.e.
$x(c)$ is $-1$ and the FDT holds. Thus for these short times the system
does not really realize that it is not in equilibrium, since the fast
degrees of freedom, such as the vibrations, are still able to follow
the dynamics of the system. This is not the case for those processes
that relax on longer time scales. In the figure we see that for times
that correspond to the aging regime, i.e. where the correlator $C$ has
fallen below the plateau, the data do no longer follow the FDT line,
but are well below it. We find that, within the accuracy of our data,
this part of the data is compatible with a straight line with slope
$-m$, with $m<1$. Such a functional form has been found in mean-field
spin glasses with one step replica symmetry breaking and thus we have
now evidence that our structural glass is compatible with this type
of spin glasses. At the moment such a connection is of course only a
tenuous one. Furthermore one might wonder whether it is really justified
to draw from a comparison of {\it nonequilibrium} properties between
systems any conclusion to {\it equilibrium} properties.  Surprisingly
for spin glasses this conclusion has been shown to be correct, see,
e.g., reference~\cite{franz98}, and thus it is not completely crazy
to assume that a similar connection can be done for structural glasses
as well. At the moment there are therefore strong efforts to test these
connections since they would allow us to gain a much more unified picture
of disordered systems. If this connection is shown to be present we would
have learned that there is no fundamental difference between disordered
systems in which the disorder is {\it quenched}, such as in spin glasses,
or in which it is {\it self-generated}, such as in structural glasses.

A different important effort to extend our understanding of aging
systems is work along the lines which have been so successful for the
equilibrium dynamics. For temperatures above $T_c$ we have seen that
mode-coupling theory \index{mode-coupling theory} is able to describe
the dynamics of supercooled liquids not only qualitatively, but also
quantitatively. Thus it is natural to try to extend this approach also
to the out-of equilibrium case since only then certain non-universal
features can be discussed on a quantitative basis.  Although in such
calculations one is faced with formidable technical problems, some
progress has recently been made~\cite{latz00} and thus it can be hoped
that in the not too far future we will have also a quantitative theory
for the aging systems.

\vspace*{5mm}

{\bf Acknowledgments}: I thank H.~C. Andersen, J.-L.~Barrat, K.~Binder,
T.~Gleim, and M.~Nauroth for the fruitful collaborations that led
to some of the results discussed in this review, and L.~Cugliandolo,
W.~G\"otze, J.~Kurchan, A.~Latz, G.~Parisi, and F.~Sciortino for many
insightful discussions on various aspects of this work. Part of this
work was supported by the SFB~261 of the DFG and the Pole Scientifique
de Mod\'elisation Num\'erique at ENS-Lyon.

\end{document}